\documentclass[12pt]{article}
\usepackage{cite,epsfig,amssymb,amsmath,graphicx,color}
\newcommand{\hs}{\hspace{-0.4cm}}

\def\be{\begin{equation}}
\def\ee{\end{equation}}
\def\ba{\begin{eqnarray}}
\def\ea{\end{eqnarray}}

\allowdisplaybreaks

\evensidemargin \oddsidemargin

\newcommand{\nn}{\nonumber}

\def\nn{\nonumber}

\begin{document}

\begin{titlepage}

\renewcommand{\thefootnote}{\fnsymbol{footnote}}


\vspace{15mm}
\baselineskip 9mm
\begin{center}
  {\Large \bf Composite-particle Hydrodynamics \\from Dyonic Black Branes }
\end{center}

\baselineskip 6mm
\vspace{10mm}
\begin{center}
 Kyung Kiu Kim,$^{1,2}$ Nakwoo Kim,$^2$ Yun-Long Zhang,$^{3,4}$
 \\[10mm]
  $^1${\sl Department of Physics and
Photon Science, School of Physics and Chemistry, Gwangju Institute
of Science and Technology, Gwangju 500-712, Korea}
  \\[3mm]
  $^2${\sl Department of Physics and
Research Institute of Basic Science, Kyung Hee University, Seoul
130-701, Korea}
     \\[3mm]
  $^3${\sl State Key Laboratory of
Theoretical Physics, Institute of Theoretical Physics, Chinese
Academy of Sciences, Beijing 100190, China}
     \\[3mm]
  $^4${\sl Mathematical Sciences and STAG Research Centre,
University of Southampton, UK}
     \\[10mm]    
  {\tt kimkyungkiu@gmail.com,~nkim@khu.ac.kr,~zhangyl@itp.ac.cn}
\end{center}

\thispagestyle{empty}

\vfill
\begin{center}
{\bf Abstract}

\end{center}
\noindent
We construct effective hydrodynamics for composite particles in
(2+1) dimensions carrying a magnetic flux by employing a holographic
approach. The hydrodynamics is obtained by perturbation of the
dyonic black brane solutions in the derivative expansion. We
introduce a consistent way to avoid mixing of different orders in
the expansion. Thanks to this method, it is possible to take the
strong external magnetic field limit in the dual field theory. To
compare our result with those for a composite particle system, we
study several cases that correspond to special solutions of
Einstein's equation and Maxwell's equations.
\\ [10mm]
~

\end{titlepage}

\baselineskip 6.6mm
\renewcommand{\thefootnote}{\arabic{footnote}}
\setcounter{footnote}{0}

\tableofcontents

\newpage

\section{Introduction}

The Anti-de Sitter/Conformal field theory(AdS/CFT) correspondence
tells us that strongly-coupled CFT's can be described by
supergravity or string theory in higher-dimensional AdS space
Ref.\cite{ads/cft}. This nontrivial relation provides many possible
techniques to study  strongly coupled field theories, which are very
difficult to investigate using conventional field theory methods.
For this reason, the AdS/CFT relation has been applied to diverse
topics: for instance supersymmetric quiver gauge theories, quantum
chromodynamics, condensed matter theory etc. In such cases, the
AdS/CFT may easily tackle the region of strong coupling, high
density or finite temperature. In this work, we use AdS/CFT to
consider another interesting region, the strong magnetic field
limit.

When an external magnetic field is applied to an electron system in
2+1 dimensions, the spectrum is given by the Landau levels.
Including interactions between the electrons, one naturally expects
the quantum Hall effect in the system. Although it is a very
non-trivial quantum mechanical phenomenon, a nice tool has been
proposed to figure out the effect. The tool is a quasi-particle
excitation around the nontrivial vacuum. The appearance of
quasi-particles such as phonons, magnons, and Cooper pair is
commonplace in condensed matter physics. For the physics under a
strong magnetic field the appropriate quasi-particle is the
composite fermion, which we find to be more advantageous for
describing the dynamics. This composite fermion is a very useful
concept to study to quantum Hall system in particular, just as
Cooper pairs for describing the mechanism of superconductors
Ref.\cite{jain89}. A nice review on composite fermions can be found
in Ref.\cite{jain}. See \textit{e.g.} Ref.\cite{jaingeneralreview}
for a more thorough exposition.

When the filling fraction $\tilde \nu =  1/2p~ $ is smaller than
1\cite{footnote1}, all the electrons occupy the lowest Landau level,
and it is the energy level that describes most of the system. Also,
the mixing with other levels can be ignored in a sufficiently strong
magnetic field. As we describe  the system using the  Landau level,
the kinetic energy is a mere constant. This means that the only
relevant part of the Hamiltonian is the interaction energy between
electrons. In order to understand the system, one should diagonalize
the Hamiltonian, which, in general, is not easy. To make matter
worse, because the only scale in the problem is given by the
interaction energy, a perturbative analysis is not available.
However, there is an interesting proposal that can help us
understand the system. It has been known for some time that some
wave-functions conjectured with the help of  experimental data
describe the system very well. They have a common factor, which
implies a situation in which every electron sees vortices at
positions occupied by other electrons. This tells us that the
effective particle of the system is the composite fermion, which is
a bound state of an electron and ``even"$(= 2p)$ quantum vortices.

Intuitively, this idea provides a nice way to understand the physics
under a strong magnetic field. Near the half-filling case, the
vortices contained in each electron cancel out the external field,
and the resulting magnetic field must be absent or very small. In
this case, we may treat the system as a collection of composite
fermions with a small external magnetic field and not in terms of
Landau levels, which behave as the composite-fermion's Fermi
sea\cite{jain89}. Following this reasoning, the particles contain
not only electric charge but also magnetic flux. This additional
ingredient should manifest itself when we calculate physical
quantities. In this paper, we show the effect in the current and the
energy-momentum tensor of hydrodynamics through a holographic
approach.

This point of view has already been exploited in the holographic
approach in Ref.\cite{BakRey}, where the authors identified a
composite fermion system with a dyonic black brane in $(3+1)$
dimensional anti-de Sitter spacetime. Some basic quantum properties
on both sides of the duality are studied. The magnetic charge of the
black brane is assumed to be proportional to its electric charge,
and the ratio between them is given by the filling fraction $\tilde
\nu$. This is the starting point for our gravity dual.

On the other hand, another important ingredient of this paper is the
fluid/gravity correspondence, which was suggested in
Ref.\cite{minwalla01}. The authors derived the boundary relativistic
hydrodynamics from the bulk Einstein's equations with a negative
cosmological constant\cite{footnote2}. There have been many
generalizations of this work to different dimensions, different
black holes, different boundary conditions and different models
including higher curvature terms\cite{raamsdonk,Battacharyya08,erdmenger08,banerjee08,KyungKiu09,Yee09,cai12,zhang12,Hu11, park11}. For the problem of our interest here, a quantum Hall
fluid, we need to consider the dyonic black holes as in
Ref.\cite{BakRey}. One is naturally led to expect that the dyonic
black hole could be dual to the magnetohydrodynamics. This duality
relation has been considered in Ref.\cite{hartnoll07,herzog07,vazquez08,Kraus2,buchbinder09,caldarelli08,hansen09,Ling12,Rene2012}.

For the study of holographic hydrodynamics, we start with the
Einstein-Maxwell action with a negative cosmological constant in
$3+1$ dimensions. Using the AdS/CFT dictionary, the dyonic black
brane solution describes a finite-temperature conformal field theory
with finite a charge density in an external magnetic field.
Macroscopic thermodynamic quantities, such as temperature (Hawking
temperature), charge density (electric charge) and magnetic field
(magnetic charge), appear as the parameters of the black brane. We
also consider a boost parameter $\vec{\beta}$ and the thermodynamic
quantities as functions of the boundary coordinates. Performing the
derivative expansion order by order, one can find a new fluctuating
black-brane solution. We consider only the leading order result here
and construct the boundary energy momentum tensor and current when
an external magnetic field is present. The expression contains the
transport coefficients of the fluid. In addition to these boundary
tensors, the bulk Einstein's equation and Maxwell's equation provide
additional constraints on the tensors. These equations play the
roles of the relativistic Navier-Stokes equations at the boundary.
When they are combined, our result describes the
magnetohydrodynamics in (2+1)-dimensional spacetime.

Adopting the approach of the fluid/gravity correspondence for dyonic
black branes, one should be careful of a perturbation in the
magnetic field. Because the interpretation of the magnetic field is
an external field in dual field theory, one cannot perturb it
without a physically-plausible reason. In our case, we do noy want
to introduce further dynamical degrees of freedom, so the magnetic
part should be kept as the original constant field. In other words,
the boosted magnetic field is written in terms of a constant
velocity field and a constant magnetic charge, and these should be
replaced by functions of the boundary coordinates in the next step
of the fluid/gravity approach. It seems that the external field is
taken as a dynamical field. To avoid this situation, we add a  field
$\delta F_{\mu\nu}$,which is given by the difference between the
original  constant field and the fluctuation field. In conclusion,
we do not touch the external field in the dual field theory, so our
computation is safe from the inconsistent situation.

In addition, there is another technical problem.  We may choose  a
gauge for the magnetic field, where the gauge potential is linear in
the boundary coordinate, $x^\mu$, for example,
$A^{\text{magnetic}}_\mu \sim  \bar H \epsilon_{\mu\nu\lambda} x^\mu
\bar u^\lambda$. Following the method of the fluild/gravity
correspondence, one can easily see that the $x^\mu$ can cause some
trouble for the expansion.  It causes mixing of  different
derivative orders and makes the magnetic field  linear in the
boundary coordinate. Thus, it is difficult to deal with a strong
magnetic field. This is one of the reasons most works using the
fluid/gravity approach concentrate on the weak-magnetic-field case.
We try to avoid this problem by  taking into account the field
strength instead of the gauge potential; then, some part of field
strength has a constraint looks like a conservation equation. This,
however, does not mean the external field is dynamical in the
boundary field theory because the total field strength is still
constant. Thus, the constraint is nothing but an identity,
$dF_{ext}=0$. We will discuss this in more detail.

Following this method, we obtained the general solution that is
consistent with the derivative expansion and the
zeroth-order(strong) magnetic field. This bulk solution is dual to
the magentohydrodynamics. As in Ref.\cite{BakRey}, we substituted
$Q(x)/\tilde \nu$ for  the magnetic field $H(x)$; then, we provided
the transport coefficients which depend on the filling fraction.  In
this case, the result resembles the magnetohydrodynamics with a
first-order (weak) external field. We point out that this effective
description is holographic dual to the transformation from the
particle system in a strong magnetic field to the composite particle
system in a weak magnetic field. In addition, we consider an example
in which the effective external magnetic field vanishes.

This paper is organized as follows. In Section II, we discuss the
derivation of hydrodynamics from the derivative expansion of
Einstein's equation. In Section III, the first-order solution is
investigated for various cases. Then, we discuss the
composite-particle case in Section \ref{sec4}  We summarize our
result and conclude in Section \ref{sec5}


\section{General structure of the metric and the field strength}
\label{sec2}

Our goal is to study a strongly coupled (2+1)-dimensional CFT system
with a charge density in a strong magnetic field. Thus, we introduce
a model which has a Maxwell's field and Einstein-Hilbert action with
a negative cosmological constant as follows:
\begin{eqnarray}\label{action}
&& \hspace{-1cm} S = \frac{1}{16\pi G} \int_\mathcal{M} d^4 x
\sqrt{-g}\left(R + \frac{6}{l^2}\right)+ \frac{1}{8\pi
G}\int_{\partial \mathcal{M}} d^3 x \sqrt{-\gamma}\Theta \nn \\
&&\hs - \frac{1}{4g_c^2}\int_{\cal M} d^4 x \sqrt{-g} \mathcal{F}^2
 + I_C~,
\end{eqnarray}
where $\Theta$ is the trace of the extrinsic curvature, the so
called Gibbons-Hawking term, and $I_C$ is a counter term, which
cancels the divergence of the action. We can write down the
expression for our configuration explicitly as follows.
\begin{eqnarray}\label{counter}
I_C = \frac{1}{8\pi G}\int_{\partial \mathcal M}d^3 x
\frac{2}{l}\sqrt{-\gamma} \left( 1 + \frac{l^2}{4}
R(\gamma_{\mu\nu})  \right)~,
\end{eqnarray}
where $\gamma_{\mu\nu}$ is the induced metric which is defined in
Eq.~(\ref{A.2}). These counter terms were introduced in
Ref.\cite{skenderis 98,kraus}, and the general results with matter
fields can be found in Ref.\cite{skenderis 00,skenderis 05,skenderis
02}. From the above action, one can write the equations of motion as
follows.
\begin{eqnarray}
&&\hs\hs R_{IJ}- \frac{1}{2}g_{IJ}R -\frac{3}{l^2} g_{IJ} - \frac{16\pi G}{2 g_c^2}({\mathcal{F}}_{KI} {{\mathcal{F}}^{K}}_J - \frac{1}{4}g_{IJ} {\mathcal{F}}^2 )\nn \\&&= 0,\\
&&\hs\hs \nabla_J {{\mathcal{F}}^{J}}_{I}=0~.
\end{eqnarray}
For convenience's sake, we define the functions related to the
equations of motion:
\begin{eqnarray}
\hs W_{IJ} &\equiv& R_{IJ} +\frac{3}{l^2} g_{IJ} - \frac{16\pi G}{2 g^2}({\mathcal{F}}_{KI} {{\mathcal{F}}^{K}}_J - \frac{1}{4}g_{IJ} {\mathcal{F}}^2 ),\\
\hs W_I &\equiv& \nabla_J {{\mathcal{F}}^{J}}_{I}~.
\end{eqnarray}
Thus, we may express the equations of motion as $W_I =0$ and
$W_{IJ}=0$.

Now, we would like to introduce an external magnetic field in the
(2+1)-dimensional system. For this, it is natural to consider the
dyonic black-brane solution in the AdS space as a gravity dual. The
dyonic uniform black-brane solution solving the equations of motion
is as follows:
\begin{eqnarray}
&&ds^2 = - r^2 f(r)dv^2 + 2  dv dr + r^2(dx_1^2 +dx_2^2
),\\\nonumber &&f(r)=   (1- \frac{\bar M}{r^3} +\frac{\bar Q^2 +
\bar H^2}{4 r^4}) , \\\nonumber &&{\mathcal{F}} = -\frac{ \bar
Q}{r^2}dr \wedge dv  +   \bar H dx_1 \wedge dx_2 ~,~\\\nonumber &&A=
\frac{\bar Q}{r} dv + \left( \int^x {dx'}^i \epsilon_{ij} \frac{
\bar H}{2} \right){dx}^j ,
\end{eqnarray}
where $i$ indices are running in $(x_1,x_2)$. We have rescaled
${\mathcal{F}}_{IJ}$ by $\frac{16 \pi G}{g_c^2}$ and took $l=1$. The
metric function $f(r)$ has two zeros, the outer horizon and the
inner horizon which are denoted by $r_+$ and $r_-$,
respectively\cite{footnote3}. Following the approach in
Ref.\cite{minwalla01}, we   consider a boosted solution:
\begin{eqnarray}\label{boosted constant background}
&&\hs ds^2 = - r^2 f(r)( \bar u_\mu dx^\mu )^2 - 2 \bar u_\mu dx^\mu
dr \nn \\ &&\quad + r^2P_{\mu \nu} dx^\mu dx^\nu,
\\\nonumber&&\hs {\mathcal{F}} =  -  \frac{\bar Q}{r^2} \bar u_\mu dx^\mu \wedge dr  - \frac{1}{2}\bar H \bar u^\lambda \epsilon_{\lambda\mu\nu} dx^\mu \wedge dx^\nu,\nonumber\\\nonumber
&&\hs \bar u^0 = \frac{1}{ \sqrt{1 - \bar \beta_i^2} },~\bar u^i =
\frac{\bar \beta_i}{ \sqrt{1 - \bar \beta_i^2}  },~P_{\mu \nu}=
\eta_{\mu\nu} + \bar u_\mu \bar u_\nu,
\end{eqnarray}
where we chose the orientation $\epsilon^{012}=1$ and $\bar u^\mu$
is a timelike vector from the boost parameter $\vec {\bar \beta}$.
In Ref.\cite{BakRey}, the authors considered $\bar H = \bar Q/\tilde
\nu $ as a holographic dual of a composite fermion system. We will
discuss that case later on.

As a next step, we perturb this solution while keeping the
macroscopic thermodynamic structure in the dual field theory, so we
regard the mass $\bar M$, the electric charge $\bar Q$, the magnetic
charge $\bar H$ and the boost parameter $\vec {\bar \beta}$ as
functions of the boundary coordinate $x^\mu$, which are denoted by
$M(x), Q(x), H(x)$ and $\vec \beta (x)$, respectively. In the dual
field theory, they are interpreted as the energy, charge density,
external magnetic field and fluid velocity, respectively. For the
field strength, except for the magnetic part, one may choose the
gauge field ansatz $ -  \frac{Q(x)}{r}u_\mu(x) dx^\mu$ , which
produces the same field strength when all functions are constant. It
is, however, not easy to find a regular gauge field ansatz for the
magnetic part. Accordingly, we had better deal with field strength
instead of the gauge field for the magnetic part. One way is to
consider ${\mathcal{F}}_{\text{mag}}= - \frac{1}{2}H(x) u^\lambda
(x) \epsilon_{\lambda\mu\nu} dx^\mu \wedge dx^\nu$ as the magnetic
part by imposing $d{\mathcal{F}}_{\text{mag}} =0$. This setup is
equivalent to taking the field strength and the metric as follows:
\begin{eqnarray}\label{rnboost}
 ds^2 &=& - r^2 f(r)( u_\mu dx^\mu )^2 - 2 u_\mu dx^\mu dr \nn
\\&&+ r^2P_{\mu \nu}(x) dx^\mu dx^\nu,
\end{eqnarray}
\begin{eqnarray}
&&\hs { \mathcal{\tilde F}} = -  \frac{Q}{r^2} u_\mu dx^\mu \wedge
dr -  \frac{\partial_\mu (Q u_\nu )}{r}dx^\mu \wedge dx^\nu \nn \\&&
\quad- \frac{1}{2}H u^\lambda \epsilon_{\lambda\mu\nu} dx^\mu \wedge
dx^\nu\label{Fstart}~~,
\end{eqnarray}
where $H(x) u^{\mu}(x)$ satisfies $\partial_\mu \left( H u^\mu \right) = 0$ due to the Bianchi identity. 
For constant $Q, H$ and $u^\mu$, the field strength becomes the
previous solution in Eq.~(\ref{boosted constant background}).  As we
explained in the introduction,  we considered this field strength
with a constraint instead of a gauge field like $A_\mu =  -\bar H
\bar u^\lambda \epsilon_{\lambda\mu\nu} x^\nu$, which produced a
singular field strength after $\bar H$ and $\bar u^\mu$ with $H(x)$
and $u^\mu(x)$.

However, this ansatz causes another problem. Unlike  $u^\mu(x)$, the
magnetic flux $H(x)$ is not a dynamical field.  Therefore, we need
to compensate for the variation of $H(x)$ by using an additional
field that does not appear in the zeroth order.  We denote such a
field as $\delta F_{\mu\nu}$, which is defined by
 \begin{eqnarray}
 \delta F_{\mu\nu} (x)&\equiv&  \partial_\mu \delta A_\nu (x)- \partial_\nu \delta A_\mu (x)\nn \\&=& -\frac{1}{2}\bar H \bar u^\lambda \epsilon_{\lambda\mu\nu}  + \frac{1}{2} H(x) u^\lambda (x) \epsilon_{\lambda\mu\nu}~~.
 \end{eqnarray}
If this is added to Eq.~(\ref{Fstart}), our starting field strength
becomes
 \begin{eqnarray}
{\mathcal{F}} &=& -  \frac{Q}{r^2} u_\mu dx^\mu \wedge dr -
\frac{\partial_\mu (Q u_\nu )}{r}dx^\mu \wedge dx^\nu \nn \\&&\hs~~-
\frac{1}{2}H u^\lambda \epsilon_{\lambda\mu\nu} dx^\mu \wedge dx^\nu
+ \frac{1}{2}\delta F_{\mu\nu} dx^\mu \wedge dx^\nu,\label{Fsstart}
 \end{eqnarray}
where the sum of the last two terms is nothing but the original
constant magnetic field; thus, we clearly do not perturb the
magnetic part and do not introduce any further dynamical degrees of
freedom.  The advantage of this compensation is that we may consider
a strong magnetic field that is consistent with the derivative
expansion.

Now, we are ready for another important step  of
Ref.\cite{minwalla01}. Because the parameters are not constant, the
metric in Eq.(\ref{rnboost}) and the field strength in
Eq.(\ref{Fsstart}) do not satisfy the equations of motion anymore.
Accordingly, $W_{IJ}$ and $W_I$ are not equal to zero, but are
proportional to the derivatives of these  functions. We call them
the source terms and denote them by $-S^{(1)}_{I}$ and
$-S_{IJ}^{(1)}~$. To obtain the first-order solution, we may add
first-order terms to the metric and the gauge field with following
ansatz \cite{footnote4}:
\begin{eqnarray}
{ds^{(n)}}^2 &=& \frac{ k_{(n)}(r)}{r}dv^2 + 2  h_{(n)} (r)dv dr + 2 r^2 j^i_{(n)}(r)dv dx^i \nn \\
&&+ r^2 (\alpha^{(n)}_{ij} - h_{(n)}(r)\delta_{ij})dx^i dx^j,  \nonumber\\
A^{(n)} &=& a^{(n)}_v (r) dv + a^{(n)}_i (r)dx^i,
\label{correction}
\end{eqnarray}
where we used ${}^{(n)}$ instead of ${}^{(1)}$ for general order,
because this ansatz is also available in for higher orders. Plugging
the above terms into $W_{IJ}$ and $W_I$, one can get additional
terms. We call them ``correction terms'', which are denoted by
$C^{(1)}_{IJ}$  and $C^{(1)}_{I}$.  Using the source terms and the
correction terms, the first-order Einstein's equation and the
Maxwell's equation can be expressed in a simple form as $
S^{(1)}_{IJ} = C^{(1)}_{IJ} $ and $S^{(1)}_I = C^{(1)}_I$.

In general, if we know the $(n-1)$th order solution, then we can
easily evaluate $S^{(n)}_{IJ}$ and $S^{(n)}_I$ ,  $C^{(n)}_{IJ}$ and
$C^{(n)}_I$ \cite{footnote5}. Therefore, the $n$th-order equations
of motion are given by
\begin{eqnarray}
&&W^{(n)}_{IJ} = C_{IJ}^{(n)} - S^{(n)}_{IJ}=0,\\
&&W^{(n)}_I = C_{I}^{(n)} - S^{(n)}_{I}=0,
\end{eqnarray}
where the equations are differential equations. The unknown
functions in Eq.(\ref{correction}) can be obtained by solving the
above equations. Not all of the correction terms, however, are
independent. For example, $C^{(n)}_v$ and $C^{(n)}_{r}$ are the same
up to some factor. The relation among such components is given by
\begin{eqnarray}
&&C^{(n)}_v + r^2 f(r) C^{(n)}_r = 0,\\
&&C^{(n)}_{vv} + r^2 f(r) C^{(n)}_{vr} = 0,  \\
&&\left\{ r^2 \left( r^2 f(r) C^{(n)}_{ri} + C^{(n)}_{vi} \right)
\right\}' + \frac{H}{2 g} \epsilon_{ij} C^{(n)}_j=0.
\end{eqnarray}
The above identities restrict the source terms as follows:
\begin{eqnarray}
&&\hs \left( S_v^{(n)} + r^2 f(r) S_r^{(n)} \right)=
0\label{constaint01},\\&&\hs
 \left(S_{vv}^{(n)} + r^2 f(r) S_{vr}^{(n)} \right)= 0 \label{constaint02},\\
 \label{constaint04}&&\hs r^4 f(r) S_{ri}^{(n)}(r) + r^2 S_{vi}^{(n)}(r)  + \int^{r}_{r_+} \frac{H}{2 } \epsilon_{ij} S_j^{(n)}(r') dr' \nn \\
&&\quad =\frac{H}{2} (Q \epsilon_{ik}  + H\delta_{ik})j^{(n)}_k
(r_+).
\end{eqnarray}
These constraint equations provide the hydrodynamics equations (the
relativistic Navier-Stokes equations).


\section{First-order hydrodynamics from gravity}
\label{sec3}

Let us compute the first-order solution explicitly. In fact, the
computation is performed in a suitable frame where the $\vec
\beta(0)$ vanishes at the origin. After all the computation in this
frame, one can easily recover the corresponding covariant form. This
clever method was suggested in Ref.\cite{minwalla01} for the first
time. To see the details of the computation, we recommend visiting
Appendix \ref{A.1}. Using the zeroth-order solution in
Eq.~(\ref{rnboost}) and (\ref{Fsstart})  and derivative expansion,
one can find the first-order source terms as listed in Appendix
\ref{A.1.2}. Putting these source terms into the constraint in
Eq.~(\ref{constaint01}) gives
\begin{eqnarray}
\partial_v Q + Q \partial_i \beta_i = 0.\label{eq3.1}
\end{eqnarray}
Changing this equation to a covariant form, we obtain a conservation
equation for the current:
\begin{eqnarray}
\partial_\mu (Q u^\mu) =0.
\end{eqnarray}
Next, the constraint in Eq.~(\ref{constaint02}) gives
\begin{eqnarray}
&&(2\partial_v M + 3 M \partial_i \beta_i)
-\frac{4}{r}[H(\partial_v H + H\partial_i \beta_i)\nn \\
&&\qquad + Q(\partial_v Q + Q \partial_i \beta_i)]= 0 .\label{dual
current conservation}
\end{eqnarray}
 Because $M$,$H$ and $\beta_i$ depend on only the boundary coordinates $x^\mu$, this is equivalent to
\begin{eqnarray}
&& H(\partial_v H + H\partial_i \beta_i)+ Q( \partial_v Q + Q
\partial_i \beta_i)=0, \label{eq3.2}
\end{eqnarray}
\begin{eqnarray}
2\partial_v M + 3 M
\partial_i \beta_i = 0. \label{eq3.3}
\end{eqnarray}
By Eq.~(\ref{eq3.1}), the first constraint in Eq.~(\ref{eq3.2}) is
nothing but another equation $\partial_\mu (H u^\mu)=0$, which has
already appeared as an assumption in the zeroth-order expression in
Eq.~(\ref{Fsstart}). Thus, this is not just an assumption but a
result of the equation of motion. The last constraint, in
Eq.~(\ref{constaint04}), is more complicated in form:
\begin{eqnarray}\label{eq3.4}
\partial_i M + 3 M \partial_v \beta_i &=& Q \delta F_{vi}-H\epsilon^{ij} \delta F_{vj} + H\epsilon^{ij}(Q \delta_{jk} -H \epsilon_{jk}  )j_k (r_+)\nonumber\nonumber\\&&  -\frac{H}{r_+}\epsilon^{ij}\left[\epsilon_{jk}\partial_k H -\delta_{jk}\partial_k Q +\left(H\epsilon_{jk} - Q \delta_{jk}   \right)\partial_v \beta_k \right].
\end{eqnarray}
The covariant forms of Eqs.~(\ref{eq3.3}) and (\ref{eq3.4}) give the
equations for the magnetohydrodynamics. We will discuss this soon.

After the calculations in Appendix \ref{A.1.3}, and up to
undetermined integration constants $\mathbb{C}_3$ and
$\mathbb{D}_i$, the first-order solutions of Einstein's equations
and Maxwell's equations can be written. Firstly, the metric is given
by
\begin{eqnarray}
ds^2 &=& -r^2 f(r) (u_\mu dx^\mu)^2 - 2 u_\mu dx^\mu dr + r^2 P_{\mu\nu}dx^\mu dx^\nu+ 2 r^2 \alpha(r)\sigma_{\mu\nu}dx^\mu dx^\nu\nonumber \\
&&+ ( r \partial_\nu u^\nu +\frac{1}{4 r^2} H u_\lambda \epsilon^{\lambda \mu\nu} \delta F_{\mu\nu} - \mathbb{C}_3 \frac{Q}{2 r^2}    )(u_\mu dx^\mu)^2 -2 r^2 j^{(1)}_\mu (r) u_\nu dx^\nu dx^\mu,
\end{eqnarray}
where the definition of the traceless symmetric tensor
$\sigma_{\mu\nu}$ and $j_\mu^{(1)}$ are
\begin{eqnarray}
\sigma_{\mu\nu} &=& \frac{1}{2}P_{\mu \alpha}P_{\nu \beta}\left[ \partial^{\alpha }u^{\beta}+\partial^{\beta }u^{\alpha} - P^{\alpha\beta}(\partial \cdot u)\right],\\
j^{(1)}_\mu(r) &=&f(r)\int_{\infty}^{r} dx \frac{1}{x^4 f(x)^2} \left\{ - x^2 u^{\nu} \partial_{\nu} u_\mu+\mathbb{D}_\mu +\frac{1}{x}\left[ \frac{2 M^2 +8M r_+^3- 16 r_+^6 }{3M} u^{\nu} \partial_{\nu} u_\mu \right. \right.\nonumber\\
&&\left.\left.+\frac{M + 2 r_+^3}{3 r_+ M} (Q P_{\mu{\nu}} - H u^\lambda \epsilon_{\lambda \mu \nu}) ( \partial^\nu Q + u^\alpha {{\epsilon_\alpha}^{ \nu}}_\beta \partial^\beta H )-\frac{4r_+(M-r_+^3)}{3M}\mathbb{D}_\mu \right]\right.\nonumber\\
&&\left. -\frac{1}{2 x^2} \left[  (Q P_{\mu \nu} - H u^\lambda
\epsilon_{\lambda \mu \nu}) ( \partial^\nu Q + u^\alpha
{{\epsilon_\alpha}^{ \nu}}_\beta \partial^\beta H ) +\frac{1}{2}(Q^2
+ H^2)u^\nu \partial_\nu u_\mu  \right]  \right\}. \nonumber\\
\end{eqnarray}
Here, the $\alpha(r)$ is given in Appendix \ref{A.1.3}, and
$\mathbb{D}_\mu$ is the covariant form of $\mathbb{D}_i$, so it
satisfies $\mathbb{D}_\mu u^\mu=0$. The gauge field is obtained by
integration. The solution is
\begin{eqnarray}
A_\mu &=& \delta A_\mu  -  \frac{Q}{r}u_\mu  - \frac{1 }{2} \left( \int^x H u^\lambda \epsilon_{\lambda \nu\mu}{dx'}^\nu \right)  -\left( \frac{\mathbb{C}_3}{r} - \frac{H}{2 r^2} u^\lambda \epsilon_{\lambda\alpha\nu}\partial^\alpha u^\nu \right)u_\mu    \nonumber\\
&&-(\partial_\mu Q +u^\lambda\epsilon_{\lambda\mu\nu}\partial^\nu H + Qu^\lambda \partial_\lambda u_\mu + H u^\lambda {\epsilon_{\lambda\mu}}^\nu u^\alpha\partial_\alpha u_\nu ) \int_\infty^r \frac{dy}{y^2 f(y)}\left(\frac{1}{r_+}-\frac{1}{y}\right) \nonumber \\
&& 
+ (Q P_{\mu \nu} + H u^\lambda\epsilon_{\lambda\mu\nu})\int_\infty^r
dy \frac{j^{(1)\nu} (y)-j^{(1)\nu} (r_+)   }{y^2 f(y)} .
\end{eqnarray}
Putting these result into Eqs.~(\ref{emtensor}) and (\ref{current}),
we get the energy-momentum tensor and the current in the boundary
theory as follows.
\begin{eqnarray}
T_{\mu\nu} &=& M(\eta_{\mu\nu}+ 3 u_\mu u_\nu ) + \mathbb{D}_\mu u_\nu + \mathbb{D}_\nu u_\mu- 2 r_+^2 \sigma_{\mu\nu},\label{stress}\\
\label{currentexp}
J^\mu 
&=&Q u^\mu + \mathbb{C}_3 u^\mu + u^\lambda {{\delta F}_\lambda}^{\mu}+ \frac{1}{3M}(Q P^{\mu\nu}+ H u^\lambda {\epsilon_\lambda}^{\mu\nu}   )\mathbb{D}_\nu \nonumber \\&&\nonumber- \frac{1}{3 r_+} \left( 1 + \frac{2 r_+^3}{M} \right) \left( P^{\mu\nu} \partial_\nu Q + u^\lambda{\epsilon_\lambda}^{\mu\nu} \partial_\nu H \right)\\&&- \frac{2}{3 r_+} \left( 1 + \frac{2r_+^3}{M}\right)\left(  Q P^{\mu\nu} + H u^\lambda {\epsilon_\lambda}^{\mu\nu}  \right) u^\sigma \partial_\sigma u_\nu .
\end{eqnarray}
Using these expressions, the constraints in Eq.~(\ref{constaint01}),
(\ref{constaint02}) and (\ref{constaint04}) can be written as
\begin{eqnarray}
&& \partial_\mu {J^{(0)}}^\mu = 0 ~~,~~\partial_\mu \left(H u^\mu
\right)  =0\\\nonumber && \partial_\mu {T^{(0)}}^{\mu\nu} =
J^{(1)}_\mu {(F_{ext}^{(0)})}^{\mu\nu} +  J^{(0)}_\mu
{(F_{ext}^{(1)})}^{\mu\nu},\label{magneto equation 1}
\end{eqnarray}
where  the current and the external field\cite{footnote6}  are
defined by
\begin{eqnarray}
&&{J^{(0)}}^\mu = Q u^\mu~ ,~
{T^{(0)}}^{\mu\nu} = M (\eta^{\mu\nu} + 3 u^\mu u^\nu ),\\
&&{F_{ext}^{(0)}} = - \frac{1}{2}H u^\lambda \epsilon_{\lambda
\mu\nu} dx^\mu  \wedge dx^\nu ,\nn \\ &&{F_{ext}^{(1)}}  =
\frac{1}{2}\delta F_{\mu\nu}  dx^\mu \wedge dx^\nu.
\end{eqnarray}
To first order, they are nothing but the magnetohydrodynamics
equations with a Bianchi identity for the external field,
\begin{align}
&\partial_\mu {J}^\mu  = 0,\quad d F_{ext}   =0,\nonumber\\
&\partial_\mu {T}^{\mu\nu} = J_\mu
F_{ext}^{\mu\nu}~.\label{emconservation}
\end{align}

\subsection{Pure Electric Black Brane Case}\label{pure electric subsection}

If we consider the case without the magnetic field ($H=0$) and take
the Landau frame, the $\mathbb{C}_3$ and the $\mathbb{D}_i$ vanish.
The current and the spacial component of the energy-momentum
conservation equation turn out to be
\begin{eqnarray}
&&\hs J^0 = Q,\nn \\
&&\hs J^i = ({F^{(1)}_{ext}})_{vi} -\frac{1}{3r_+}(1 +
\frac{2r_+^3}{M})\left( \partial_i Q + 2 Q \partial_v \beta_i
\right),\\&&\hs \partial_i M + 3M \partial_v \beta_i = Q
({F^{(1)}_{ext}})_{vi} .
\end{eqnarray}
From these two equations, one can easily find different expressions
for the current which is standard in hydrodynamics,
\begin{eqnarray}
J^\mu = Q u^\mu + \sigma_1 \left(  u_\nu ({F_{ext}^{(1)}})^{\nu\mu}
- T P^{\mu\nu} \partial_\nu \frac{\mu}{T}  \right)~,
\end{eqnarray}
where $T$ and $\mu = \frac{Q}{r_+}$ are the
temperature\cite{footnote7} and the chemical potential of the dual
field theory, respectively. The electric conductivity function
$\sigma_1$ is given by
\begin{eqnarray}
\sigma_1 =  \left(  \frac{12 r_+^4 - Q^2}{3 (4 r_+^4 + Q^2)}
\right)^2. \label{conductivity 1}
\end{eqnarray}
~\\
The boundary energy momentum tensor and the hydrodynamics
equations are given by
\begin{eqnarray}
&&T_{\mu\nu}= M(\eta_{\mu\nu}+ 3 u_\mu u_\nu)-2r^2_+ \sigma_{\mu\nu},\\
&&\partial_\mu T^{\mu\nu}= J_\mu
{(F_{ext}^{(1)})}^{\mu\nu}~~,~~\partial_\mu J^\mu =0~~.
\end{eqnarray}
This result is the (2+1)-dimensional version of Ref.\cite{Kyung Kiu
09}.

\subsection{Small Velocity and Constant Magnetic Field Limit}
In Ref.\cite{Kraus2}, the authors studied a dyonic black brane with
small velocity and constant magnetic field. It is possible to cover
this limit using our result. In order to get their result, we take
${(F^{(1)}_{ext})}_{\mu\nu}=0$, $\vec \beta =0$ and constant $H$.
From the energy-momentum conservation equation,
Eq.~(\ref{emconservation}), we get the current $J_i = \frac{1}{H}
\epsilon_{ji}\partial_i M$. We already have the current expression,
Eq.~(\ref{currentexp}). In this case, the current becomes
\begin{eqnarray}
J_{i} &=& -\frac{1}{3r_+}\left( 1 + \frac{2 r_+^3}{M}  \right)
\left(\partial_i Q -\epsilon_{ij}\partial_j H  \right) \nn\\ &&+
\frac{1}{3M}(Q\delta_{ij} -H \epsilon_{ij})\mathbb{D}_j.
\end{eqnarray}
Subtracting the currents, one can find $\mathbb{D}_i$, which is just
$T_{vi}$:~\\
\begin{eqnarray}\label{krausDi}
\mathbb{D}_i = T_{vi} = \frac{3M}{Q^2 +
H^2}\left[\frac{Q}{H}\epsilon_{ij}\partial_j M - \partial_i M  +
\frac{1}{3r_+}\left( 1 + \frac{2r_+^3}{M}  \right) \left(
Q\partial_i Q + H\epsilon_{ij}\partial_j Q \right)     \right]~.
\end{eqnarray}
This is consistent with the result in Ref.\cite{Kraus2}.


\section{Fluid dynamics for composite particles}
\label{sec4}

In this section, we present the transport coefficients of the fluid
dynamics corresponding to the dyonic black brane. In what follows we
assign the filling fraction as the ratio of the charge density to
the external field as in Ref.\cite{BakRey}.  Such an identification
of the magnetic field shows how the coefficients depend on the
filling fraction. In addition,  we provide an example dual to a
special solution of the bulk equation in which the boundary current
is proportional to the Poincare dual of the external field.

\subsection{ Magnetohydrodynmics  Corresponding to a Dyonic Black Brane  }\label{more general case}

Now, we are ready to get  the magnetohydrodynamics from
Eqs.~(\ref{stress}), (\ref{currentexp}) and (\ref{emconservation}).
The conventional forms of the current and the energy-momentum tensor
do not contain $u^\sigma \partial_\sigma u_\nu$, so we have to
substitute for it. Using the conservation equation for the
energy-momentum tensor in Eq.~(\ref{emconservation}), one can
replace it with derivatives of macroscopic thermodynamics  functions
such as $\partial_\mu \mu$, $\partial_\mu T$, and so on. Taking the
Landau frame with $\mathbb{C}_3=0$  and  $\mathbb{D}_\mu=0$, the
boundary current and the energy-momentum tensor are as follows.
\begin{eqnarray}
T_{\mu\nu} &=& M(\eta_{\mu\nu} + 3 u_\mu u_\nu ) -2 r_+^2 \sigma_{\mu\nu},\\
\nonumber   J^\mu &=& Q u^\mu+ \left(\Sigma_1 P^{\mu\nu}- \Sigma_2
u_\lambda \epsilon^{\lambda\mu\nu} \right)\partial
_\nu\left(\frac{\mu }{T}\right)\nn \\ &&+\left( \Sigma_3 P^{\mu\nu}
- \Sigma_4 u_\lambda \epsilon^{\lambda\mu\nu} \right)\partial _\nu
\left(\frac{m}{T}\right)\text{    }\\&&+ \left( \Sigma_5 P^{\mu\nu}
-\Sigma_6 u_\lambda \epsilon^{\lambda\mu\nu} \right)u^\sigma \delta
F_{\sigma \nu}  ~~,\label{general current}
\end{eqnarray}
where the energy density $M$ is given by $r_+^3 +\frac{Q^2 + H^2}{4
r_+} $. In addition, $\mu=\frac{Q}{r_+}$ and $m=\frac{H}{r_+}$ are
the chemical potential and a quantity proportional to the
magnetization, respectively. The temperature $T$ and the outer
horizon $r_+$ are given by
\begin{eqnarray}
T &=& \frac{12 r_+^2 - \mu^2 -m^2}{16\pi r_+},\nn \\ r_+ &=&
\frac{4\pi T+ \sqrt{(4 \pi T)^2+ 3(\mu^2 + m^2)}}{6}.
\end{eqnarray}
The transport coefficients are presented in Appendix \ref{A.3}. The
magnetohydrodynamics equations are as follows:
\begin{eqnarray}
&&\partial_\mu T^{\mu\nu} = J_\mu
{(F_{ext})}^{\mu\nu},~~\partial_\mu J^\mu=0,\label{emcon 01}
\end{eqnarray}
where $(F_{ext})_{\mu\nu} = -H \epsilon_{\mu\nu\lambda}u^\lambda +
\delta F_{\mu\nu} =-\bar H \epsilon_{\mu\nu\lambda} \bar u^\lambda$,
the constant external field.

\subsection{Composite-particle System}\label{small}

 Because we are interested in a fluid that consists of the composite particles, we may employ the identification of the magnetic field,  $\bar H =\bar Q/{\tilde \nu}$, in Ref.\cite{BakRey} and look at the effect in the formulation. As we mentioned, $\tilde \nu$ is the filling fraction related to a Chern-Simons level in dual field theory. Keeping $\tilde \nu$ constant, one can follow the method used in the fluid/gravity approach. The result is easily obtained by substituting $Q(x)/\tilde \nu$ for $H(x)$ in the previous subsection.  Under this  substitution, the right hand side of Eq.(\ref{emcon 01}) shows an interesting feature:
\begin{align}
J_\mu {(F_{ext})}^{\mu\nu} &= {(J^{(0)})}_\mu  {(F_{ext}^{(1)})}^{\mu\nu} + {(J^{(1)})}_\mu  {(F_{ext}^{(0)})}^{\mu\nu}\nonumber\\
&= Q u_\mu \delta F^{\mu\nu} -  (J^{(1)})_\mu    \frac{1}{\tilde \nu} Q  u_\lambda \epsilon^{\lambda\mu\nu}\nonumber\\
&= Qu_\mu\left(  \delta F^{\mu\nu}  + \frac{1}{\tilde \nu}
\epsilon^{\lambda\mu\nu}   (J^{(1)})_\lambda  \right).
\end{align}
Thus, the effective hydrodynamics equation becomes
\begin{eqnarray}
\partial_\mu T^{\mu\nu} = J_\mu {(F_{\text{eff}})}^{\mu\nu},~~~ \partial_\mu J^{\mu}=0, \label{magneto effective field}
\end{eqnarray}
where ${(F _{\text{eff}})}_{\mu\nu}$ is given by $\delta F_{\mu\nu}
+ \frac{1}{{\tilde \nu}} \epsilon_{\mu\nu\lambda}
(J^{(1)})^\lambda$, which plays the role of a first-order external
field. The expressions for the current and the energy-momentum
tensor can be rearranged as follows\cite{footnote8}:
\begin{eqnarray}
\hs J^\mu &=& (Q+\mathbb{C}_3)u^\mu - \sigma_2 T
P^{\mu\nu}\partial_\nu \frac{\mu}{T} +\frac{1}{1 + 1/{\tilde
\nu}^2}\sigma_2 u_\nu (F_{\text{eff}})^{\nu\mu} \nn \\&&-
\frac{1}{{\tilde \nu}}\frac{1}{1 + 1/{\tilde \nu}^2} u_\lambda
\epsilon^{\lambda\mu\nu}u^\sigma (F_{\text{eff}})_{\sigma\nu}+
\frac{Q}{3M}\mathbb{D}^\mu, \nonumber
\\\hs T_{\mu\nu}&=&
M(\eta_{\mu\nu} + 3 u_{\mu}u_\nu) -2 r_+^2 \sigma_{\mu\nu}
+\mathbb{D}_\mu u_\nu +\mathbb{D}_\nu u_\mu,\label{current and
energy momentum eff}
\end{eqnarray}
where the conductivity $\sigma_2$ is given by
\begin{eqnarray}
\sigma_2 =\left( \frac{12 r_+^4 - (1+ 1/{\tilde \nu}^2)Q^2  }{3( 4
r_+^4 + (1+ 1/{\tilde \nu}^2)Q^2 )} \right)^2.
\end{eqnarray}
Replacing $Q$ with $Q/\sqrt{1 + 1/{\tilde \nu}^2} $, this quantity
becomes the conductivity for the pure electric case in
Eq.~(\ref{conductivity 1}). One can see that, in this
composite-particle picture, the magnetohydrodynamics with a strong
(zeroth order) magnetic field can be changed into the fluid dynamics
with a weak (first order) magnetic field as Section III.~1, except
for the Hall current part.  In addition,  one can write down the
current with ${\delta F}_{\mu\nu}$ instead of
$(F_{\text{eff}})_{\mu\nu}$ in the Landau frame as follows.
\begin{eqnarray}
T_{\mu\nu} &=& M(\eta_{\mu\nu} + 3 u_\mu u_\nu ) -2 r_+^2
\sigma_{\mu\nu},\\J^\mu &=& Q u^\mu + (\Sigma_{a} P^{\mu\nu} -
\Sigma_{b} u_\lambda \epsilon^{\lambda \mu\nu})\partial_\nu
\frac{\mu}{T} \nn \\ &&+ (\Sigma_{c} P^{\mu\nu} - \Sigma_{d}
u_\lambda \epsilon^{\lambda \mu\nu})u^\sigma \delta F_{\sigma\nu}
,\label{current ext}
\end{eqnarray}
The transport coefficients are given in Appendix \ref{A.3}. The
above current form is a special case of the more general
consideration in Ref.\cite{Rene 2012}. One can compare it to this
expression.


\subsection{Example with Vanishing $(F_{\text{eff}})_{\mu\nu}$ }

Now, we consider a special case where the effective external field
vanishes. This situation is simply realized when $\delta F_{\mu\nu}
= - \frac{1}{\tilde \nu}  \epsilon^{\lambda\mu\nu}
(J^{(1)})_\lambda$.  Incorporating this with $H= Q/\tilde \nu$, one
can write a relation between the external field and the current as
follows:
\begin{eqnarray}
J^\mu = \frac{\tilde \nu}{2} \epsilon^{\mu\nu\lambda}
(F_{ext})_{\nu\lambda}~~.\label{CS constraint}
\end{eqnarray}
It is well known that this condition is the equation of motion of a
Chern-Simons theory whose action is $S_{\text{eff}} \sim \int d^3 x
\left( \tilde \nu A_{ext}\wedge dA_{ext} + J^\mu ({A_{ext}})_\mu +
\cdots \right)$.  Plugging Eq.~(\ref{CS constraint}) into
Eq.~(\ref{current and energy momentum eff}), the current and the
momentum become
\begin{eqnarray}
&&\hs\hs J^{\mu} = Q u^\mu + \frac{\tilde
\nu}{2}\epsilon^{\mu\nu\lambda} \label{hall current 01} \delta
F_{\nu\lambda},  \\&&\hs \hs T_{\mu\nu}= M(\eta_{\mu\nu} + 3
u_{\mu}u_\nu) -2 r_+^2 \sigma_{\mu\nu} +\mathcal{D}_\mu u_\nu
+\mathcal{D}_\nu u_\mu,\label{hall energy momentum 01}
\end{eqnarray}
where $M$ and $\mathcal{D}_\mu$ are given by
\begin{eqnarray}
M&=& r_+^3 + \left( 1+\frac{1}{{\tilde \nu}^2} \right)\frac{Q^2}{4 r_+} \label{M D},  \\
\mathcal{D}^\mu &=& 
\frac{(12 r_+^4 -(1 + 1/ {{\tilde \nu}^2})Q^2)^3}{192 Q \pi r_+^4 (4
r_+^4 +(1 + 1/{{\tilde \nu}^2})Q^2)} P^{\mu\nu}\partial_\nu
\frac{\mu}{T}\nn \\ &&-\frac{3{\tilde \nu} (4 r_+^4 +(1 + 1/{{\tilde
\nu}^2})Q^2)}{4 r_+ Q}u_\nu \epsilon^{\mu\nu\lambda}
 u^\sigma \delta F_{\sigma\lambda}
.\nonumber\\
\end{eqnarray}
The spatial part of the current in Ref.~(\ref{hall current 01})
gives the exact Hall conductivity $\sigma_{xy} =  {\tilde \nu}$, and
the temporal component gives the charge density corrected by the
Chern-Simons constraint as follows:
\begin{eqnarray}
J^0 = Qu^0 + {\tilde \nu} \delta F_{12} =\tilde \nu \bar H u^0 ~~.
\end{eqnarray}
The energy-momentum tensor provides a non-vanishing $T_{0i}$
component. One can see that the second term of $\mathcal D_\mu$
plays the role of the Hall momentum. Also, the magnetohydrodynamics
equation, Eq.~(\ref{emconservation}), is changed as follows.
\begin{eqnarray}\label{strong hydro eq}
\partial_\mu J^\mu =0 ~~,~~ \partial_\mu T^{\mu\nu}=0 ~.
\end{eqnarray}

As we intended, the  energy momentum tensor is divergence free. This
reminds us of the well-known fact discussed in the introduction. One
advantage of the composite-fermion paradigm is that one can
transform the electron system with an external magnetic field into a
composite-particle system without external magnetic field or with a
small magnetic field. Such a transformation can be described by a
Chern-Simons theory with an external field. Since we are dealing
with the same situation descried by Eq.~(\ref{CS constraint}),we can
conclude that our result is a holographic version for the
transformation\cite{jain general review}.


\section{Discussion}
\label{sec5}

In this paper, we have investigated a strongly-coupled field-theory
system in an external magnetic field. Using the holographic
approach, this system is realized by using dyonic black branes. The
electric and the magnetic charges of the dyonic branes correspond to
the charge density and the external field in dual-field theory,
respectively. Because we consider a (3+1)-dimensional AdS spacetime,
the dual system is (2+1)-dimensional, and this bulk solution could
be dual to a quantum Hall system.

The important property and the interesting phenomena of the quantum
Hall system are captured by transport coefficients, which can be
measured in laboratories. This measurement is nicely encoded in the
hydrodynamics, a purely-phenomenological effective model. In
addition, it is well known that the hydrodynamics for a Super
Yang-Mills fluid can be derived from the long-wavelength limit of
fluctuating black holes\cite{minwalla01}. This is one of the
successful result in AdS/CFT. We have followed the method and
derived the hydrodynamics for the dyonic black brane.

In the fluid/gravity correspondence approach, the
magnetohydrodynamics type equation was studied for the first time in
Ref.\cite{Kyung Kiu 09}. The authors considered a Reissner-Nordstrom
black brane in (4+1)-dimensional AdS spacetime. They started with a
boundary condition on the bulk gauge field, $(A_{ext})_\mu = \lim_{r
\rightarrow \infty} A_\mu(r)$, which has no contribution to the
zeroth-order solution.

A boundary coordinate dependence in the external field like
$(A_{ext})_\mu(x)$ makes a correction to the hydrodynamics equation.
Finally one can obtain the magnetohydrodynamics. The external field,
however, is given by $(F_{ext})_{\mu\nu} =\partial_\mu
(A_{ext})_\nu-\partial_\nu (A_{ext})_\mu$, and this field strength
is first order in the derivative expansion, \textit{i.e.}, the
weak-external-field limit. Thus, our work is devoted to introducing
the zeroth-order or strong external field. In order to realize such
a configuration, we chose the field strength in Eq.~(\ref{boosted
constant background}).  For the constant $\bar H$ and $\bar u_\mu$,
it is easy to find a corresponding gauge potential. While one takes
them as functions of the boundary coordinates $x^\mu$, the existence
of a gauge potential requires $\partial_\mu ( H u^\mu)=0$  in the
first order. As we pointed out in the introduction and Section III,
this has nothing to do with a dynamical current. It is just part of
the trivial equation $d F_{ext} =0$, and  it is consistent with one
of the equations of motion Eq.~(\ref{dual current conservation}), so
we do not need further constraint.

In order to check our result with other works, we discussed some
examples. Our first example were a pure electric black brane without
a magnetic field. It was presented in the subsection III.~1. This
first order RN black brane solution is dual to a fluid with a
conserved current. We obtained the electric conductivity in the
(2+1)-dimensional system. This is the (2+1)-dimensional version of
Ref.\cite{Kyung Kiu 09}. Also, the second example is the energy
momentum tensor in Ref.\cite{Kraus2}. In their paper, they
considered a configuration where the fluid velocity vanished and the
zeroth-order external field $H$ was constant. Our result reproduced
their first order result.

In this work, we tried to relate the composite fermion or particle
system to the dyonic black brane by using the fluid/gravity
approach. This duality was conjectured in Ref.\cite{BakRey}, where
the authors considered a magnetic field, $Q/\tilde \nu $.  In our
consideration, we obtained the magnetohydrodynamics equation,
Eq.~(\ref{magneto effective field}), with the effective external
field $(F_{\text{eff}})^{\mu\nu}=\delta F^{\mu\nu} + \frac{1}{\tilde
\nu} \epsilon^{\mu\nu\lambda}(J^{(1)})_\lambda$, where
$(J^{(1)})_\lambda$ is the first-order part of the current. The
current and the energy-momentum tensor are given in
Eq.~(\ref{current and energy momentum eff}). The expression shows
that the current is very similar to the pure black-brane electric
case except for the Hall current in the Landau frame. Also, the
electric conductivity has the same structure, with different
electric charge $Q_{\text{eff}}= \sqrt{1 + 1/\tilde \nu^2} Q$.

To see the simplest case, we investigated the situation where
$F_{\text{eff}}$ vanished. Such a condition is equivalent to
Eq.~(\ref{CS constraint}), which has the same form as the equation
of motion for a Chern-Simons theory with the filling fraction
$\tilde \nu$. By construction, this describes a quantum Hall fluid
exactly, and the Hall conductivity $\tilde \nu$ can be read from
Eq.~(\ref{hall current 01}).  In addition, we have observed that
there exists a Hall-current-type momentum flow in the
energy-momentum tensor, Eq.~(\ref{hall energy momentum 01}). In this
example, we would like to point out that the hydrodynamics equation
is a magnetohydrodynamics-type equation like $\partial_\mu
T^{\mu\nu} =J_\mu (F_{ext})^{\mu\nu}$  in general; however the RHS
vanishes when the fluid satisfies the equation of motion in
Eq.~(\ref{CS constraint}) for the Chern-Simons  theory. This is the
same situation that appears in the quantum Hall system. To
understand this special situation, one needs reminding that the
composite fermions are bound states of a charged fermion and
magnetic fluxes. Using this quasi-particle as the basic degrees of
freedom, the external magnetic field can be effectively canceled by
the fluxes with some fixed filling fraction. This means that an
electron system in strong magnetic field can be described by a
composite-fermion system without a magnetic field. We realized this
cancelation holographically by showing that the final hydrodynamics
equations were the conservation equation without an external field,
\textit{i.e.}, $\partial_\mu T^{\mu\nu}=0$.

As future directions, it could be valuable to study the interesting
duality transformation among the transport coefficients in
Ref.\cite{BakRey}, but we did not include it at this moment. In
order to consider the duality, one has to include an axion term
$\int \theta F \wedge F $ in our action Ref.\cite{BakRey} and it
produces an important change in the fluid/gravity correspondence
approach. In addition, another interesting problem is the role of
$\int \mathcal{R}\wedge \mathcal{R}$ in the magnetohydrodynamics. In
the boundary field theory, this is related to the Hall viscosity and
the angular momentum density, which are another important topics in
hydrodynamics\cite{Saremi:2011ab,Liu:2012zm}.

\appendix
\section{Appendix}

\subsection{Solving equations of motion}
\label{A.1}

\subsubsection{First-order source terms}
\label{A.1.2} In order to find the first order solution, we need to
obtain the first-order source terms. We follow the clever way
suggested in Ref.\cite{minwalla01}. Using the field strength and the
metric in Eq.~(\ref{rnboost}), we can calculate the first-order
source terms. Because there is a plane symmetry in the boundary, an
efficient way for this calculation is to take a special position in
the plane. The most convenient choice is the origin, which does not
cause any loss in generality. In addition, we may take an
appropriate frame where $\beta^i(0)=0$\cite{footnote9}; then, the
calculation in this frame becomes much simpler. Taking a derivative
expansion of Eq.~(\ref{rnboost}) and putting it into $W_{IJ}$ and
$W_I$, one can calculate the source terms from the scalar and the
tensor parts of Einstein's equations as follows.
\begin{eqnarray}
&&S^{(1)}_{rr} = 0,\\
&&S^{(1)}_{rv} =
\frac{ H \epsilon^{ij} Q \partial_i \beta_j }{2 r^5}-\frac{H \epsilon^{ij}\delta F_{ij} }{4 r^4}+\frac{\partial_i \beta_i }{r},\\
&&S^{(1)}_{vv} = r \partial_v f(r) + \frac{H \epsilon^{ij} \delta F_{ij}    f(r)  }{4 r^2} - \frac{ \left( r^2 f(r) \right)' \partial_i \beta_i }{2}- \frac{  H \epsilon^{ij} Q \partial_i \beta_j      f(r)}{2r^3},\\
&&S^{(1)}_{ij} =\left\{\frac{H \epsilon^{kl} \delta F_{kl}}{4 r^2} -
\frac{Q H \epsilon^{kl}\partial_k \beta_l }{2r^3} + r\partial_i
\beta_i \right\} \delta_{ij} + r (\partial_i \beta_j+\partial_j
\beta_i),
\end{eqnarray}
Also, the source terms for the vector parts of Einstein's equations
and Maxwell's equations are given by
\begin{eqnarray}
S^{(1)}_{vi} &=&
\frac{4r^4 + 2 r M + Q^2 -H^2  }{4 r^3} \partial_v \beta_i -\frac{ H\epsilon_{ij}\partial_j Q+ H \epsilon_{ij}Q \partial_v \beta_j - r \partial_i M + H\partial_i H }{2 r^3}\\\nonumber &&+ \frac{ H\epsilon_{ij}\delta F_{vj} - \delta F_{vi} Q }{2 r^2},\\
S^{(1)}_{ri} &=&-\frac{\partial_v \beta^i }{r},\\
S^{(1)}_{i} &=&-  \frac{\partial_i Q - \epsilon_{ij}\partial_j H + Q
\partial_v \beta_i - H \epsilon_{ij}\partial_v \beta_j }{ r^2}~.
\end{eqnarray}
In addition, the other scalar parts of Maxwell's equations are given
as follows.
\begin{eqnarray}
&&S^{(1)}_{r} = -   \frac{H \epsilon^{ij}\partial_i \beta_j   }{r^4},\\
&&S^{(1)}_{v} =  \frac{ \partial_v Q + Q \partial_i \beta_i }{r^2}+
\frac{H \epsilon^{ij}\partial_i \beta_j f(r)}{ r^2}~.
\end{eqnarray}

\subsubsection{General correction terms }
\label{A.1.1} Putting Eq.~(\ref{correction}) into $W_{IJ}$ and
$W_I$, we can evaluate $C^{(n)}_{IJ}$ and $C^{(n)}_I$, which are
given by
\begin{eqnarray}
C^{(n)}_{vv} &=& f(r) \left\{-\left(6 r^2+\frac{H^2}{r^2}\right) h^{(n)}(r)-r^2 \left(r^2 f(r)\right)' {h^{(n)}}'(r)-\frac{r k^{(n)\prime\prime }(r)}{2}+\frac{Q}{2  } {a^{(n)}_v}'(r)\right\}\nonumber,\\
&&\\
C^{(n)}_{vr} &=& \left(6+\frac{H^2}{r^4}\right) h^{(n)}(r)+\left(r^2 f(r)\right)' {h^{(n)}}'(r)+\frac{{k^{(n)}}''(r)}{2 r}-\frac{Q}{2   r^2} {a^{(n)}_v}'(r),\\
C^{(n)}_{rr} &=& \frac{1}{r^4} \left\{r^4 {h^{(n)}}'(r)\right\}',\\
C^{(n)}_{vi} &=& \frac{H}{2 r^2} \left(H \delta _{i j}+Q \epsilon _{i j}\right) j^{(n)}_j(r)-f(r) \left\{\frac{1}{2} \left(r^4 {j^{(n)}_i}'(r)\right)'-\frac{Q}{2 } {a^{(n)}_i}'(r)\right\},\\
C^{(n)}_{ri}&=& \frac{1}{2 r^2} \left(r^4 {j^{(n)}_i}'(r)\right)'-\frac{Q \delta _{i j}+H \epsilon _{i j}}{2   r^2} {a^{(n)}_j}'(r),\\
C^{(n)}_{ij} &=& -\frac{\left\{r^4 f(r) {\alpha^{(n)} _{i j}}'(r)\right\}'}{2}\nonumber\\
&&+\delta _{i j} \left\{\left(6 r^2-\frac{H^2}{r^2}\right)
h^{(n)}(r)+\frac{\left(r^8 f(r) {h^{(n)}}'(r)\right)'}{2
r^4}+{k^{(n)}}'(r)+\frac{Q}{2  } {a^{(n)}_v}'(r)\right\},
\end{eqnarray}
and
\begin{eqnarray}
C^{(n)}_v &=& f(r) \left\{r^2 {a^{(n)}_v}'(r)+2   Q h^{(n)}(r)\right\}',\\
C^{(n)}_r &=& -\frac{1}{r^2} \left\{r^2 {a^{(n)}_v}'(r)+2   Q h^{(n)}(r)\right\}',\\
C^{(n)}_i &=&  \left\{r^2 f(r) {a^{(n)}_i}'(r)-  \left(Q \delta _{i
j}-H \epsilon _{i j}\right) j^{(n)}_j(r)\right\}',
\end{eqnarray}
where we have taken $l=1$.

\subsubsection{First-order solutions}
\label{A.1.3}

Given the source terms in Appedix \ref{A.1.2}, one can find the
correction terms in Eq.~(\ref{correction}) in the metric and the
gauge field by simple integrations. The scalar and  the tensor parts
are related to the source terms $S^{(1)}_{rr}$, $ S^{(1)}_r$,
$\delta^{ij}S^{(1)}_{ij}$ and $ S^{(1)}_{ij} -
\frac{1}{2}\delta_{ij}S^{(1)}_{kk} $. For these combinations, we
obtained
\begin{eqnarray}
&&h^{(1)}(r) = \mathbb{C}_1 + \frac{ \mathbb{C}_2}{r^3}, \\
&&a^{(1)}_v(r) = \frac{ H}{2 r^2} \epsilon^{ij}\partial_i \beta_j +\mathbb{C}_2 \frac{  Q}{2 r^4} + \frac{\mathbb{C}_3}{r} + \mathbb{C}_4,   \\
&&k^{(1)}(r)= -\frac{1}{4r} H \epsilon^{ij} \delta F_{ij} + r^2 \partial_i \beta_i,\nonumber  \\
&&~~~~~~~~~~~~ -  \mathbb{C}_1(2 r^3 + \frac{H^2}{r} )   +   \mathbb{C}_2 ( -\frac{Q^2}{4 r^4 } + \frac{M}{2 r^3} -\frac{H^2}{4 r^4}   ) -  \mathbb{C}_3 \frac{Q}{2   r} + \mathbb{C}_5,  \\
&&\alpha^{(1)}_{ij}(r) \equiv \alpha(r)\left(\partial_i \beta_j +
\partial_j \beta_i -\delta_{ij}(\partial_k \beta^k \right),
\end{eqnarray}
where $\alpha(r)$ and its asymptotic behavior are given by
\begin{eqnarray} \alpha(r) = - \int_{\infty}^{r}\left( \frac{y^2 -
r_+^2}{y^4 f(y)}  \right)dy = \frac{1}{r}-\frac{r_+^2}{3
r^3}+\frac{M}{4 r^4}-\frac{Q^2+H^2}{20
r^5}+O\left(\frac{1}{r^6}\right).
\end{eqnarray}
In the above result, we have many integration constants. Because the
first integration constant $\mathbb{C}_1$ will give a
non-normalizable mode, this should vanish. For $\mathbb{C}_2$, one
can always set this to zero by using a coordinate transformation of
$r$. $\mathbb{C}_4$ can be taken as zero because this is pure gauge.
A non-vanishing $\mathbb{C}_5$ means a non-vanishing trace of the
energy-momentum tensor; this must vanish because of the conformal
symmetry. Thus, the only remaining constant is $\mathbb{C}_3$.

Considering the vector part equations $W_{ri}$ and $W_i$, we got a
second-order differential equation as follows:
\begin{eqnarray}
r^2 f(r)\left( r^2 {j_i^{(1)}} (r)\right)'' - r^2 j^{(1)}_i (r)
\left( r^2 f (r)\right)'' = \frac{1}{r^2} \zeta_i (r)~~,
\end{eqnarray}
where $\zeta_i(r)$ is given by
\begin{eqnarray}
\zeta_i(r) &=& 2r^4  f(r) S_{ri}^{(1)}(r) -(H^2 + Q^2)j_i^{(1)}(r_+)
+  (Q \delta_{ij} + H \epsilon_{ij})\int_{r_+}^r dx
S_j^{(1)}(r)\nonumber\\\nonumber&=&-2r^3 f(r) \partial_v\beta_i -
(H^2 + Q^2)j_i^{(1)}(r_+)\nonumber \\ &&+ (\frac{1}{r} -
\frac{1}{r_+}) (Q\delta_{ij} + H \epsilon_{ij})\left\{ \partial_j Q
-\epsilon_{jk}\partial_k H + Q \partial_v \beta_j - H
\epsilon_{jk}\partial_v \beta_k   \right\} .
\end{eqnarray}
The general solution is
\begin{eqnarray}
j_i^{(1)}(r)=\mathbb{D}_{1i} f(r) + f(r) \int_{r_1}^r dx
\frac{1}{x^4 f(x)^2} \left\{ \int_{r_0}^x dy \frac{ \zeta_i
(y)}{y^2} + \mathbb{D}_{2i}   \right\}~~.
\end{eqnarray}
The $r=\infty$ boundary condition gives $\mathbb{D}_{1i} = 0$ and
$r_1 = \infty$: then, the solution becomes
\begin{eqnarray}
 j_i^{(1)}(r) =  f(r) \int_\infty^r  dx \frac{1}{x^4 f(x)^2} \left\{ \Delta_i (x) +  \mathbb{D}_{i} \right\}~~,
\end{eqnarray}
where $\Delta_i (r)$ has a slightly complicated form:
\begin{eqnarray}
\Delta_i (r) &=& -  r^2 \partial_v \beta_i+\frac{1}{r} (H^2 +
Q^2)j_i^{(1)}(r_+) \nonumber \\\nonumber \nonumber
&&+ \frac{1}{r}\left[\frac{1}{r_+}(Q\delta_{ij} + H \epsilon_{ij})( \partial_j Q -\epsilon_{jk}\partial_k H)+2(M-2r_+^3)\partial_v\beta_i     \right]\\
&&-\frac{1}{2r^2}\left[(Q\delta_{ij} + H \epsilon_{ij})(\partial_j Q
-\epsilon_{jk}\partial_k H )+(H^2 + Q^2)\partial_v \beta_i\right].
\end{eqnarray}
When we compute the boundary tensors, \textit{i.e.}, the
energy-momentum tensor and the current, those are given by the
asymptotic forms of the metric and the Maxwell field. Thus, one has
to look at the asymptotic form of $j_i^{(1)}$, which is as follows
\begin{eqnarray}
&&j_i^{(1)}(r)=  \frac{\partial_v \beta_i}{r} - \frac{ \mathbb{D}_i
}{3r^3}-\frac{(H^2 + Q^2)}{4r^2}j_i^{(1)}(r_+)\\\nonumber
&&\qquad\quad~-\frac{1}{4r^2r_+}(Q\delta_{ij} + H \epsilon_{ij})(\partial_j Q -\epsilon_{jk}\partial_k H + Q \partial_v \beta_j - H \epsilon_{jk}\partial_v \beta_k)
+O\left(\frac{1}{r^5}\right).
\end{eqnarray}
From the integration of the $W_i$ source term, we have the gauge
field
\begin{eqnarray}
a_i^{(1)}(r) =    (Q \delta_{ij} -H \epsilon_{ij})\int_{\infty}^r dx
\frac{j_j^{(1)}(x)- j_j^{(1)}(r_+) }{x^2 f(x)} + \int_{\infty}^r
\frac{dy}{y^2 f(y)} \int_{r_+}^y dx S_i^{(1)}(x),
\end{eqnarray}
where $S^{(1)}_{i} = -\frac{\partial_i Q - \epsilon_{ij}\partial_j H
+ Q \partial_v \beta_i - H \epsilon_{ij}\partial_v \beta_j }{ r^2}$
and we have taken the boundary condition  $a_\mu^{(1)} (r=\infty)=
0$. The asymptotic form of the gauge field near the boundary is
\begin{eqnarray}
a_i^{(1)}(r) 
&=& \frac{1}{r}\left[(Q \delta_{ij} -H \epsilon_{ij})j_j^{(1)} (r_+) + \frac{ 1}{r_+}( \partial_i Q - \epsilon_{ij}\partial_j H + Q \partial_v \beta_i - H \epsilon_{ij}\partial_v \beta_j ) \right] +O\left(\frac{1}{r^2} \right).\nonumber\\
\end{eqnarray}
 Using the asymptotic expressions, we can read the boundary tensors $J^\mu$ and $T^{\mu\nu}$, which are defined in the section \ref{A.2}.

In order to find $j_i^{(1)}(r_+)$, we should consider regularity
conditions for the metric. The metric should be smooth at the
horizon, so $\mathbb{D}_i$ can be determined by using
\begin{eqnarray}
\mathbb{D}_i =   -3 M j_i^{(1)}(r_+) +\frac{ n_i}{r_+^2}+3
\partial_v \beta_i r_+^2~~,
\end{eqnarray}
where $n_i$ is
\begin{eqnarray}
&&n_i = -\frac{1}{2}(Q\delta_{ij} + H \epsilon_{ij})\left\{
\partial_j Q -\epsilon_{jk}\partial_k H   \right\} - \frac{H^2 +
Q^2}{4}\partial_v \beta_i~~.
\end{eqnarray}
One can check that the metric is smooth at the horizon by using this
$\mathbb{D}_i$. Thus, $j_i^{(1)}(r_+)$ can be expressed in terms of
$\mathbb{D}_i$:
\begin{eqnarray}
j_i^{(1)}(r_+) = \frac{ n_i}{3 M r_+^2}+ \frac{r_+^2}{M}  \partial_v
\beta_i  -\frac{\mathbb{D}_i}{3M}~~.
\end{eqnarray}

\subsection{Boundary stress energy tensor and current }
\label{A.2} In this section, we briefly describe the prescription
for the boundary stress-energy tensor and current. We take the
prescription from Refs.\cite{skenderis 98, kraus, skenderis
00,skenderis 05}. Our metric can be expressed in the following form
under the Arnowitt-Deser-Misnet(ADM) decomposition:
\begin{equation}
ds^2 = \gamma_{\mu\nu}(dx^\mu + V^\mu dr)(dx^\nu + V^\nu dr) + N^2
dr^2~~.
\end{equation}
Using this expression, one can obtain the stress energy tensor by
varying the on-shell action with respect to the induced metric. The
resulting boundary stress-energy tensor is given by
\begin{equation}\label{emtensor}
  T_{\mu\nu} \equiv \lim_{r \rightarrow \infty }  r \frac{-2}{\sqrt{-\gamma}} \frac{\delta S_{cl}}{\delta \gamma^{\mu\nu}}   =\lim_{r \rightarrow \infty }  r  [  -2 ( \Theta_{\mu\nu} - \Theta \gamma_{\mu\nu} + \frac{2}{l}\gamma_{\mu\nu}- l G_{\mu\nu} ) ]~~,
\end{equation}
where the last two terms came from the counter term in
Eq.(\ref{counter}) and $S_{cl}$ is the on-shell action. Here, the
extrinsic curvature $\Theta_{\mu\nu}$ is
\begin{equation}
\Theta_{\mu\nu} = \frac{1}{2 N} [\partial_r \gamma_{\mu\nu} - D_\mu
V_\nu - D_\nu V_\mu ]~~.
\end{equation}
In addition, the boundary current can be obtained by varying the
action with respect to the gauge field as follows:
\begin{eqnarray}\label{current}
J^\mu &=& \lim_{r \rightarrow \infty} r^3 \frac{-1}{\sqrt{-\gamma}}
\frac{\delta S_{cl}}{\delta \tilde A_\mu} =  \lim_{r \rightarrow
\infty} r^3 N F^{r \mu} \\\nonumber &=& \lim_{r \rightarrow \infty}
r^3 \frac{1}{N} [ \gamma^{\mu\lambda}( A'_\lambda  - \partial
_\lambda A_r ) - V^\lambda \gamma^{\mu\sigma}(\partial_\lambda
A_\sigma - \partial_\sigma A_\lambda)  ],
\end{eqnarray}
where $\tilde A_\mu$ is the gauge field projected to the boundary.

\subsection{Transport coefficients}
\label{A.3} The coefficients in Eq.(\ref{general current}) are given
by
\begin{eqnarray}
\Sigma_1  &=&\Sigma^{-1}_0 \left[(1+\sigma_G)\sigma_A-\sigma_H\sigma_B \right],\nonumber\\
\Sigma_2  &=&\Sigma^{-1}_0 \left[(1+\sigma_G)\sigma_B+\sigma_H\sigma_A \right],\nonumber\\
\Sigma_3  &=&\Sigma^{-1}_0 \left[(1+\sigma_G)\sigma_C-\sigma_H\sigma_D \right],\nonumber\\
\Sigma_4  &=&\Sigma^{-1}_0 \left[(1+\sigma_G)\sigma_D+\sigma_H \sigma_C\right],\nonumber\\
\Sigma_5  &=&\Sigma^{-1}_0 \left[(1+\sigma_G)(1+\sigma_E)-\sigma_H\sigma_F \right],\nonumber\\
\Sigma_6  &=&\Sigma^{-1}_0 \left[(1+\sigma_G)\sigma_F+\sigma_H(1+\sigma_E) \right],
\end{eqnarray}
where $\Sigma_0 = (1+\sigma_G)^2+\sigma_H^2$ and
\begin{eqnarray}
\sigma_A&=&+\frac{Q}{3 r_+}\left(1+\frac{2 r_+^3}{M}\right)\left(\frac{\mu}{T}\sigma_0-\frac{T}{\mu}\right),\quad \sigma_E=-\frac{2Q^2}{9Mr_+}\left(1+\frac{2 r_+^3}{M}\right),\nonumber\\
\sigma_B&=&-\frac{H}{3 r_+}\left(1+\frac{2 r_+^3}{M}\right)\frac{\mu}{T}\sigma_0,\qquad\qquad\quad \sigma_F=+\frac{2H Q}{9Mr_+}\left(1+\frac{2 r_+^3}{M}\right),\nonumber\\
\sigma_C&=&+\frac{Q}{3 r_+}\left(1+\frac{2
r_+^3}{M}\right)\frac{m}{T}\sigma_0,\qquad\qquad\quad
\sigma_G=-\frac{2H^2}{9Mr_+}\left(1+\frac{2 r_+^3}{M}\right),\nonumber\\
\sigma_D&=&+\frac{H}{3 r_+}\left(1+\frac{2 r_+^3}{M}\right)\left(\frac{m}{T}\sigma_0-\frac{T}{m}\right),\quad \sigma_H=+\frac{2HQ}{9Mr_+}\left(1+\frac{2 r_+^3}{M}\right),
\end{eqnarray}
as well as
\begin{eqnarray}
\sigma_0=\frac{2T^2(36 r_+^4+H^2+Q^2)r_+^2}{2(4
r_+^4+H^2+Q^2)(12r_+^4+H^2+Q^2)}~~.
\end{eqnarray}
Also, the coefficients in Eq.(\ref{current ext}) are obtained
through
\begin{eqnarray}
\Sigma_a  &=&\Sigma_1+\Sigma_3/\tilde \nu,\qquad\Sigma_c  =\Sigma_5,\nonumber\\
\Sigma_b  &=&\Sigma_2+\Sigma_4/\tilde \nu,\qquad\Sigma_d  =\Sigma_6,
\end{eqnarray}
where we also have used identification $H\equiv Q/\tilde \nu$.

~\\ 
{\bf Ackowledgments}

K.K. thanks Kimyeong Lee and Sang-Jin Sin for helpful discussions on
the early setup of this work and for useful comments. This work was
supported by a National Research Foundation of Korea (NRF) grant
funded by the Korea government(MEST) with grant No.~2010-0023121,
2011-0023230, 2012046278, and also through the Center for Quantum
Spacetime (CQUeST) of Sogang University (grant number 2005-0049409).
In addition, this work was supported by a Korea Institute for
Advanced Study(KIAS) grant funded by the Korea government(MSIP).
Y.L.Z. is thankful for the helpful comments from R. G. Cai, M. M.
Caldarelli and K. Skenderis, as well as the support by China
Scholarship Council (No.~201204910341) and National Natural Science
Foundation of China (No.~10821504, No.~10975168 and No.~11035008).
The work of KKK was supported by Basic Science Research Program
through the National Research Foundation of Korea (NRF) funded by
the Ministry of Science, ICT \& Future Planing
(NRF-2014R1A1A1003220).


\begin{thebibliography}{99}









 \bibitem{ads/cft}
J.~M.~Maldacena, Int. J. Theor. Phys. \textbf{38}, 1113 (1999);
Adv.\ Theor.\ Math.\ Phys.\ {\bf 2}, 231 (1998)
  [arXiv:9711200 [hep-th]].


\bibitem{jain89}
 J.~K.~Jain,
  Phys.\ Rev.\ Lett.\  {\bf 63}, 199 (1989).




\bibitem{jain}
J.~K.~Jain, 
 Phys. Today {\bf 53}, 39 (2000).


\bibitem{jaingeneralreview}
For a general review of composite fermion physics, see J.~K.~Jain,
 Adv. Phys. {\bf 41}, 105 (1992) and references therin.


\bibitem{footnote1}
{Each electron possesses $2p$ magnetic flux quanta, where $p$ is an
integer.  From now on we use $\tilde \nu$  instead of $1/2p$.  }


\bibitem{BakRey}
  D.~Bak and S-J.~Rey,
  JHEP {\bf 1009}, 032 (2010)
  [arXiv:0912.0939 [hep-th]].


\bibitem{minwalla01}
  S.~Bhattacharyya, V.~E.~Hubeny, S.~Minwalla and M.~Rangamani,
  JHEP {\bf 0802}, 045 (2008)
  [arXiv:0712.2456 [hep-th]].

\bibitem{footnote2}
{Actually this result is believed to have wider applicability than
AdS/CFT, because we may start with any gravitational system with a
well-defined thermodynamics.}



\bibitem{raamsdonk}
  M.~Van Raamsdonk,
  JHEP {\bf 0805}, 106 (2008)
  [arXiv:0802.3224 [hep-th]].




\bibitem{Battacharyya08}
  S.~Bhattacharyya, R.~Loganayagam, S.~Minwalla, S.~Nampuri, S.~P.~Trivedi and S.~R.~Wadia,
  JHEP {\bf 0902}, 018 (2009)
  [arXiv:0806.0006 [hep-th]].

\bibitem{erdmenger08}
  J.~Erdmenger, M.~Haack, M.~Kaminski and A.~Yarom,
  JHEP {\bf 0901}, 055 (2009)
  [arXiv:0809.2488 [hep-th]].

\bibitem{banerjee08}
  N.~Banerjee, J.~Bhattacharya, S.~Bhattacharyya, S.~Dutta, R.~Loganayagam and P.~Surowka,
  JHEP {\bf 1101}, 094 (2011)
  [arXiv:0809.2596 [hep-th]].

\bibitem{KyungKiu09}
  J.~Hur, K.~K.~Kim and S-J.~Sin,
  JHEP {\bf 0903}, 036 (2009)
  [arXiv:0809.4541 [hep-th]].

\bibitem{Yee09}
  M.~Torabian and H-U.~Yee,
  JHEP {\bf 0908}, 020 (2009)
  [arXiv:0903.4894 [hep-th]].


\bibitem{cai12}
  R-G.~Cai, L.~Li, Z-Y.~Nie and Y-L.~Zhang,
  Nucl.\ Phys.\ B {\bf 864}, 260 (2012)
  [arXiv:1202.4091 [hep-th]].

\bibitem{zhang12}
  X.~Bai, Y-P.~Hu, B-H.~Lee and Y-L.~Zhang,
  JHEP {\bf 1211}, 054 (2012)
  [arXiv:1207.5309 [hep-th]].


\bibitem{Hu11}
Y. Hu, H. Li , Z. Nie, JHEP {\bf 01}, 123 (2011), [arXiv:1012.0174
[hep-th]].

\bibitem{park11}
  Y-P.~Hu and C.~Park,
  Phys.\ Lett.\ B {\bf 714}, 324 (2012)
  [arXiv:1112.4227 [hep-th]].


\bibitem{hartnoll07}
  S.~A.~Hartnoll, P.~K.~Kovtun, M.~Muller and S.~Sachdev,
  Phys.\ Rev.\ B {\bf 76}, 144502 (2007)
  [arXiv:0706.3215 [cond-mat.str-el]].

\bibitem{herzog07}
  S.~A.~Hartnoll and C.~P.~Herzog,
  Phys.\ Rev.\ D {\bf 76}, 106012 (2007)
  [arXiv:0706.3228 [hep-th]].

\bibitem{vazquez08}
  E.~I.~Buchbinder, S.~E.~Vazquez and A.~Buchel,
  JHEP {\bf 0812}, 090 (2008)
  [arXiv:0810.4094 [hep-th]].

\bibitem{Kraus2}
  J.~Hansen and P.~Kraus,
  JHEP {\bf 0904}, 048 (2009)
  [arXiv:0811.3468 [hep-th]].

\bibitem{buchbinder09}
  E.~I.~Buchbinder and A.~Buchel,
  Phys.\ Rev.\ D {\bf 79}, 046006 (2009)
  [arXiv:0811.4325 [hep-th]].

\bibitem{caldarelli08}
  M.~M.~Caldarelli, O.~J.~C.~Dias and D.~Klemm,
  JHEP {\bf 0903}, 025 (2009)
  [arXiv:0812.0801 [hep-th]].

\bibitem{hansen09}
  J.~Hansen and P.~Kraus,
  JHEP {\bf 0910}, 047 (2009)
  [arXiv:0907.2739 [hep-th]].

\bibitem{Ling12}
  C-Y.~Zhang, Y.~Ling, C.~Niu, Y.~Tian and X-N.~Wu,
  Phys.\ Rev.\ D {\bf 86}, 084043 (2012)
  [arXiv:1204.0959 [hep-th]].


\bibitem{Rene2012}
  K.~Jensen, M.~Kaminski, P.~Kovtun, R.~Meyer, A.~Ritz and A.~Yarom,
  JHEP {\bf 1205}, 102 (2012)
  [arXiv:1112.4498 [hep-th]].


\bibitem{skenderis98}
  M.~Henningson and K.~Skenderis,
  JHEP {\bf 9807}, 023 (1998)
  [hep-th/9806087].
  M.~Henningson and K.~Skenderis,
  Fortsch.\ Phys.\  {\bf 48}, 125 (2000)
  [hep-th/9812032].

\bibitem{kraus}
  V.~Balasubramanian and P.~Kraus,
  Commun.\ Math.\ Phys.\  {\bf 208}, 413 (1999)
  [hep-th/9902121].
\bibitem{skenderis00}
  S.~de Haro, S.~N.~Solodukhin and K.~Skenderis,
  Commun.\ Math.\ Phys.\  {\bf 217}, 595 (2001)
  [hep-th/0002230].
\bibitem{skenderis05}
  I.~Papadimitriou and K.~Skenderis,
  JHEP {\bf 0508}, 004 (2005)
  [hep-th/0505190].



\bibitem{skenderis02}
  K.~Skenderis,
  Class.\ Quant.\ Grav.\  {\bf 19}, 5849 (2002)
  [hep-th/0209067].



\bibitem{footnote3}
{ The relations with the black hole parameters are given by $\bar
M=r_-^3+r_+ r_-^2+r_+^2 r_-+r_+^3$ and $\bar Q^2+ \bar H^2 =4 r_-
r_+ \left(r_-^2+r_+ r_-+r_+^2\right)$. }




\bibitem{footnote4}
{Where $\delta^{ij}\alpha_{ij} = 0$, and we took a gauge choice,
$g_{rr}=0,~g_{r\mu} \propto u_\mu,~Tr[ {(g^{(0)})}^{-1} g^{(n)} ]=0,
~\forall  n> 0$. In addition, we take a convenient frame which can
be always recovered to the covariant form with $ u_\mu $,
$P_{\mu\nu}$ by the boost. }

\bibitem{footnote5}
{ $C^{(n)}_{IJ}$ and $C^{(n)}_I$  are listed in Appendix
\ref{A.1.1}. }


\bibitem{footnote6}
{   More explicitly, $(F_{ext})_{\mu\nu} = (F_{ext}^{(0)})_{\mu\nu}
+  (F_{ext}^{(1)})_{\mu\nu} = -H(x) u^\lambda  (x)
\epsilon_{\lambda\mu\nu}  + \delta F_{\mu\nu} (x)= - \bar H \bar
u^\lambda \epsilon_{\lambda\mu\nu} . ~$ }


\bibitem{footnote7}
{This is given by the Hawking temperature in the bulk solution by
AdS/CFT. }



\bibitem{footnote8}
{$\mathbb{C}_3$ and $\mathbb{D}_\mu$ vanish in the Landau frame and
$M$ is given by $r_+^3 + \left( 1+\frac{1}{{\tilde \nu}^2} \right)
\frac{Q^2}{4 r_+}$ .}


\bibitem{Saremi:2011ab}
  O.~Saremi and D.~T.~Son,
  JHEP {\bf 1204}, 091 (2012)
  [arXiv:1103.4851 [hep-th]].

\bibitem{Liu:2012zm}
  H.~Liu, H.~Ooguri, B.~Stoica and N.~Yunes,
  Phys.\ Rev.\ Lett. {\bf 110}, 211601 (2013)
  [arXiv:1212.3666 [hep-th]].









\bibitem{footnote9}
{One should notice that $\partial^n \beta^i|_{x=0}$ does not vanish
in generic situation.}






























\end{thebibliography}
\end{document}